\documentclass[conference]{styles/IEEEtran}
\newcommand{\PaperPreprint}{}
\IEEEoverridecommandlockouts

\usepackage{styles/infocom2027}
\begin{document}

\title{CoSlice: Coordinated Multi-Agent Reinforcement Learning for 5G Network Slicing Allocation}

\ifdefined\PaperPreprint
\author{
\IEEEauthorblockN{Pengxin Wang\IEEEauthorrefmark{1}, Nakjung Choi\IEEEauthorrefmark{2}, Jingdi Chen\IEEEauthorrefmark{1}}
\IEEEauthorblockA{\IEEEauthorrefmark{1}Department of Electrical and Computer Engineering, University of Arizona}
\IEEEauthorblockA{\IEEEauthorrefmark{2}Nokia \& Bell Labs}
}

\else
\author{
\IEEEauthorblockN{Anonymous Authors}
}
\fi

\maketitle

\begin{abstract}
Radio Access Network (RAN) slicing is a shared resource allocation problem: each slice seeks to satisfy its Service Level Agreement (SLA) with minimal radio-resource consumption, while all slices compete for a limited resource budget. This creates both an intra-slice trade-off between SLA assurance and resource efficiency and an inter-slice conflict over shared resources, calling for specialized yet coordinated control. In this work, we present CoSlice, a coordinated multi-agent reinforcement learning framework for inter-slice physical resource block allocation in Open Radio Access Network (O-RAN). CoSlice uses preference-conditioned slice agents for fast resource allocation and an LLM-based coordinator to adapt their objectives using global network observations at a slower timescale. We formulate the problem as a multi-objective partially observable stochastic game (mo-POSG), establish the existence of a Preference Correlated Equilibrium, and derive a sublinear regret bound for the learning procedure. We implement the algorithm by using a two-stage training procedure that consists of multi-objective pretraining and joint optimization. We build a network slicing evaluation environment using ORANSlice and its radio-frequency simulator (RFsim), supporting realistic protocol-stack execution and application traffic across multiple slices. Experimental results show that CoSlice reduces both average and tail SLA violations compared with other centralized and multi-agent baselines under diverse slice workloads.

\end{abstract}

\begin{IEEEkeywords}
Network Slicing, Multi-Agent Multi-Objective Reinforcement Learning, Game Theory
\end{IEEEkeywords}

\section{Introduction}

\label{sec:introduction}

Network slicing is a key capability of 5G and beyond networks \cite{network_slicing_survey}, which logically partitions a physical Radio Access Network (RAN) into multiple virtual networks with heterogeneous Service-Level Agreement (SLA) requirements. Typical slice types include enhanced Mobile Broadband (eMBB), Ultra-Reliable Low-Latency Communication (URLLC), and massive Machine-Type Communication (mMTC), each targeting different throughput, latency, and reliability requirements. Managing these slices requires adaptive inter-slice radio resource allocation. In the Open Radio Access Network (O-RAN) architecture \cite{3gpp_slice_management, oran_whitepaper}, programmable xApps and rApps enable intelligent network slicing policies, making learning-based RAN slice management increasingly attractive.

Recent studies have applied deep reinforcement learning (DRL) to network slicing by enabling agents to learn adaptive control policies directly from network traces and runtime statistics~\cite{drl_slicing, drl_slicing_survey, eexapp}, where many use a centralized agent that jointly processes the traffic and QoS states of all slices. However, different slices exhibit distinct traffic patterns and SLA objectives: eMBB mainly emphasizes throughput, whereas URLLC prioritizes latency and reliability. Mixing these signals within a single policy makes slice specialization difficult. A natural alternative is to assign a dedicated agent to each slice ~\cite{COQRA,DUAC,onslicing,crl_slicing,hrl_slicing}. Each agent can focus on its own traffic, QoS state, and SLA objective. Nevertheless, all agents still compete for the same limited PRB budget. Requesting more PRBs may improve one slice's QoS, but reduces the resources available to others. Therefore, independently optimized agents may produce conflicting requests and inefficient allocations, making inter-slice coordination essential.

Another challenge of radio resource control is the stringent latency requirement~\cite{oran_whitepaper}. MAC scheduling operates at approximately 1ms time scale, leaving little time for model based inference or iterative planning. At the Near-RT RIC, where control loops typically range from 10ms to 1s, deploying large models or performing autoregressive LLM decoding can introduce prohibitive inference latency. In contrast, complex network trace data often require aggregation across slices and computationally intensive processing to support effective coordination. Network slicing must therefore reconcile fast resource decisions with slower and more computationally demanding network wide coordination.

To address these problems, we present \emph{CoSlice}, a coordinator sub-agents framework for dynamic inter-slice PRB allocation in O-RAN. CoSlice deploys one preference-conditioned agent for each slice in the Near-Real-Time control loop. Each agent observes only its local traffic and QoS conditions and maps them, together with a slice specific objective preference, to a PRB request. An LLM coordinator is deployed in Non-Real-Time RIC to analyze aggregated monitoring reports over longer time horizons and adjust the agents’ multi-objective preferences, which balances SLA satisfaction and resource efficiency and enable adaptive network-wide coordination under changing network conditions. 

The main contributions of this paper are as follows: (1) We develop a multi-agent system with preference coordination for RAN slicing, where each slice agent learns a resource request policy from its local traffic and SLA objectives, while an LLM coordinator adjusts multi objective preferences based on high-level statistics. (2) We formulate the interactions among slice agents as a preference correlated game. We prove the existence of an equilibrium and establish a sublinear regret bound under policy and value approximation errors. (3) We design a two stage algorithm that combines multi objective pretraining with multi-agent joint optimization, and evaluate it using ORAN-Slice and OAI RFsim under diverse traffic loads and resource constraints.

\section{Related Work}
\label{sec:related-work}

\noindent\textbf{Machine Learning for Inter-slice PRB Allocation.}
Recent studies have explored machine learning based method for inter-slice RAN resource allocation problem. Zipper~\cite{zipper} argues that slice level SLAs are too coarse, and uses app level throughput and latency SLAs as the optimization objective. This motivates our design to consider fine grained QoS feedback and slice granularity. A line of work ~\cite{onslicing}~\cite{adaslicing}~\cite{xslice}, highlights the need for online adaptation under dynamic network states, and updates slicing policies while preserving SLA constraints. However, online policy updates may suffer from instability and delayed adaptation. Beyond single policy optimization, COQRA~\cite{COQRA} studies inter slice coordination with multiple slice managers that exchange action value information, while DUAC~\cite{DUAC} optimizes distributional utilities for slice performance guarantee. These works motivate our design in QoS modeling, regret based reward design and multi slice coordination. Unlike most existing approaches that use constrained reinforcement learning as the formulation, which minimize PRB usage subject to SLA constraints, we model SLA violation and PRB usage as two competing objectives under a multi-objective reinforcement learning framework.

\textbf{LLM for Network Optimization.}
Recent advances in large language models (LLMs) show great potential for network control systems. As the pioneer to apply LLM to network problems, NetLLM~\cite{netllm} fine-tunes pretrained LLMs for networking tasks by enabling them to process multi-modal network data and replacing the language modeling head with task specific networking heads. LLM-hRIC~\cite{LLM_ORAN} uses an LLM in the Non-RT RIC to generate a coarse power allocation policy from global network information. This policy serves as an initial guidance for RL exploration in the Near-RT RIC and is gradually replaced by the learned RL policy during training. ORION~\cite{orion} uses an LLM to parse natural language service intents and translate them into structured slice requirements, including throughput, latency, reliability, service area, and duration. Instead of directly control network slice, we use LLM to generate preference signals for better coordination of slice sub-agents.

\section{Preliminaries and Problem Formulation}
\label{sec:preliminaries}

In this work, we formulate the network slicing problem as a multi-objective partially observable stochastic game (MO-POSG)~\cite{felten2026momaland} equipped with a coordination mechanism, defined by the tuple $(\mathcal{G},\mathcal{C})$, where $\mathcal{G}=\langle\mathcal{N},\mathcal{S},\mathcal{A},\mathcal{O},P,\Omega,\boldsymbol{\ell},\gamma\rangle$. Here, $\mathcal{N}=\{1,\dots,N\}$ is the set of slice agents, $\mathcal{S}$ is the global network state space, $\mathcal{A}=\prod_{i\in\mathcal{N}}\mathcal{A}_i$ is the joint action space, and $\mathcal{O}=\prod_{i\in\mathcal{N}}\mathcal{O}_i$ is the joint observation space. Each local action $a_i\in\mathcal{A}_i$ denotes the PRB request of slice agent $i$. The transition kernel $P$ governs the network dynamics, $\Omega$ is the observation kernel, and $\gamma\in[0,1)$ is the discount factor.

The vector $\boldsymbol{\ell}=(\ell_1,\ldots,\ell_N)$ contains the slice-level multi-objective cost functions. For each slice agent $i$, the cost function $\ell_i:\mathcal{S}\times\mathcal{A}\rightarrow\mathbb{R}_{\ge0}^{2}$ returns the cost vector $\ell_i(s_t,\mathbf{a}_t)=(\ell_i^{\mathrm{SLA}}(s_t,\mathbf{a}_t),\ell_i^{\mathrm{PRB}}(s_t,\mathbf{a}_t))^\top$. The PRB component is the normalized resource request, i.e., $\ell_i^{\mathrm{PRB}}(s_t,\mathbf{a}_t)=a_{i,t}/A_{\max}$, where $A_{\max}$ denotes the maximum available PRB budget. The SLA component measures the step-level relative SLA violation:
\begin{equation}
\label{eq:relative_sla_penalty}
\ell_i^{\mathrm{SLA}}(s_t,\mathbf{a}_t)
=
\frac{1}{|\mathcal{M}_i|}
\sum_{m\in\mathcal{M}_i}
\left[
\frac{s_m\left(x_{m,t}^i-x_{m,\mathrm{sla}}^i\right)}
{x_{m,\mathrm{sla}}^i}
\right]^+.
\end{equation}
Here, $[x]^+=\max(x,0)$, $\mathcal{M}_i$ denotes the set of SLA metrics associated with slice $i$, and $x_{m,\mathrm{sla}}^i$ is the corresponding SLA threshold for metric $m$. The direction indicator is set to $s_m=-1$ for metrics with a minimum requirement (e.g., throughput) and $s_m=+1$ for metrics with a maximum requirement (e.g., latency and packet loss).

As the coordination mechanism, $\mathcal{C}$ maps the traffic history $h_t\in\mathcal{H}$ to a collection of correlated preference vectors $\boldsymbol{z}_t=(\boldsymbol{z}_{1,t},\ldots,\boldsymbol{z}_{N,t})\in\mathcal{Z}$ according to $\boldsymbol{z}_t\sim\mathcal{C}(\cdot\mid h_t)$. For each slice agent $i$, the preference vector $\boldsymbol{z}_{i,t}=(z_{i,t}^{\mathrm{SLA}},z_{i,t}^{\mathrm{PRB}})^\top$ specifies the relative importance of SLA violation and PRB usage, resulting in the scalarized cost $\bar{\ell}_i(s_t,\mathbf{a}_t,\boldsymbol{z}_{i,t})=\langle\boldsymbol{z}_{i,t},\ell_i(s_t,\mathbf{a}_t)\rangle$. Conditioned on $\boldsymbol{z}_{i,t}$, slice agent $i$ follows the stochastic policy $\pi_i(a_i\mid o_i,\boldsymbol{z}_i)$, where $o_i\sim\Omega_i(\cdot\mid s)$ is its local observation. The corresponding action-value and state-value functions are defined as $Q_i^{\boldsymbol{\pi}}(s,\mathbf{a},\boldsymbol{z}_i)=\mathbb{E}_{\boldsymbol{\pi}}\!\left[\sum_{k=0}^{\infty}\gamma^k\bar{\ell}_i(s_{t+k},\mathbf{a}_{t+k},\boldsymbol{z}_{i,t+k})\,\middle|\,s_t=s,\mathbf{a}_t=\mathbf{a}\right]$ and $V_i^{\boldsymbol{\pi}}(s,\boldsymbol{z}_i)=\mathbb{E}_{\boldsymbol{\pi}}\!\left[\sum_{k=0}^{\infty}\gamma^k\bar{\ell}_i(s_{t+k},\mathbf{a}_{t+k},\boldsymbol{z}_{i,t+k})\,\middle|\,s_t=s\right]$, respectively, where $\boldsymbol{\pi}=(\pi_1,\ldots,\pi_N)$ denotes the joint policy.

Based on the above formulation, our goal is to learn a preference-conditioned policy profile under the coordination mechanism $\mathcal{C}$ such that no slice agent can further reduce its expected discounted scalarized cost through unilateral policy deviation.

\section{Theoretical Analysis}
\label{sec:theory}
In this section, we study the theoretical properties of the proposed preference-conditioned game. We first establish the existence of Preference Correlated Equilibrium (PCE) in Subsec.~\ref{subsec:pce}, and then analyze the convergence of the proposed learning framework toward such an equilibrium in Subsec.~\ref{subsec:pce_convergence}.

\subsection{Existence of Preference Correlated Equilibrium}
\label{subsec:pce}

To characterize the solution of the preference-conditioned Markov game
$(\mathcal{G},\mathcal{C})$, we adapt the notion of correlated equilibrium (CE)~\cite{ce}
into Preference Correlated Equilibrium (PCE). In contrast to correlated equilibrium in normal form game, where a correlation device recommends private actions $a_i$, our coordinator $\mathcal{C}$ provides latent preference signals $z_i$ that serve as conditioning variables for the agents.

Formally, a policy profile $\boldsymbol{\pi}^*=(\pi_1^*,\dots,\pi_N^*)$ is a PCE if, for every slice agent $i\in\mathcal{N}$, every global history $h\in\mathcal{H}$, every received preference signal $z_i\in\mathcal{Z}_i$, and every alternative policy $\pi_i'\in\Pi_i$,
\[
\begin{aligned}
&\mathbb{E}_{\mathbf z_{-i}}
\left[
    J_i(
        \pi_i^*,
        \boldsymbol{\pi}_{-i}^*,
        z_i,\mathbf z_{-i}
    )
\right] \\
&\le
\mathbb{E}_{\mathbf z_{-i}}
\left[
    J_i(
        \pi_i',
        \boldsymbol{\pi}_{-i}^*,
        z_i,\mathbf z_{-i}
    )
\right],
\qquad \forall \pi_i'\in\Pi_i .
\end{aligned}
\]
where $\mathbf z_{-i}=(z_1,\dots,z_{i-1},z_{i+1},\dots,z_N)$ denotes the preference signals received by all agents except $i$. The expected discounted cost of slice agent $i$ is defined as
\[
J_i(\pi_i,\boldsymbol{\pi}_{-i},z_i,\mathbf z_{-i})
=
\mathbb{E}_{\tau}
\left[
    \sum_{t=0}^{\infty}\gamma^t
    \ell_i(s_t,\mathbf a_t,z_i)
\right],
\]
where $\tau=(s_0,\mathbf o_0,\mathbf a_0,s,\dots)$ denotes the trajectory induced by the preference signals $\mathbf z=(z_i,\mathbf z_{-i})$, and the preference-conditioned policy profile $\boldsymbol{\pi}=(\pi_i,\boldsymbol{\pi}_{-i})$.

Now we establish the existence of PCE in our setting. To this end, we make the following regularity assumption: the cost functional $J_i$ is quasi-convex with respect to its own policy $\pi_i$.

\begin{theorem}
\label{theo: PCE_exist}
In our Markov Game $\mathcal{G}$ with correlated signal $\mathcal{C}$, if Assumptions mentioned above are satisfied, there exists a PCE policy profile $\boldsymbol{\pi}^*=(\pi_1^*,\dots,\pi_N^*)$ such that no agent with given signal has the desire to deviate from its policy unilaterally.
\end{theorem}

\begin{proof}[Proof of theorem \ref{theo: PCE_exist}]
We prove the existence of PCE through Glicksberg’s fixed point theorem~\cite{fixed_point_theorem}, which is commonly used for continuous games with infinite dimensional policy spaces.

\textit{Step 1: Verify Compactness and Convexity of Joint Policy space.} Let $\Pi_i$ denote the preference-conditioned policy space with $\pi_i(\cdot\mid o_i,z_i)\in\Delta(\mathcal A_i)$. In our setting, the action denotes a relaxed PRB request in $\mathcal A_i=[A_{\min},A_{\max}]$, which is compact and convex. Since $\mathcal O_i\times\mathcal Z_i$ is compact and $\Delta(\mathcal A_i)$ is compact and convex under the weak topology, the policy space $\Pi_i$ is compact and convex, and the joint policy space $\Pi=\Pi_1\times\cdots\times\Pi_N$ is also compact and convex.

\textit{Step 2: Define the Best Response Correspondence.} For each slice agent $i$, let $\Pi_{-i}=\prod_{j\ne i}\Pi_j$. The best response
correspondence $BR_i:\Pi_{-i}\rightrightarrows \Pi_i$ is defined as
\[
BR_i(\boldsymbol{\pi}_{-i})
=
\arg\min_{\pi_i\in\Pi_i}
\bar J_i(\pi_i),
\]
where $\bar J_i(\pi_i)=
\mathbb{E}_{\mathbf z_{-i}}
[
J_i(\pi_i,\boldsymbol{\pi}_{-i},z_i,\mathbf z_{-i})
]$ denotes the expected cost of agent $i$ for given $\boldsymbol{\pi}_{-i}$ and $z_i$. Furthermore, since the SLA penalty cost $\ell_i(o_i,\mathbf a)$ is continuous with respect to $o_i$ and $\mathbf{a}$, $\bar J_i(\pi_i)$ is continuous with respect to $\pi_i$. With compactness of $\Pi_i$, the minimum is attainable, so $BR_i(\boldsymbol{\pi}_{-i})$ is well-defined.

\textit{Step 3: Verification of the conditions for Glicksberg's fixed point theorem.} We now verify that the best response correspondence satisfies the required conditions.

\textit{Compact and Convex Valuedness.} Consider any convergent sequence $\{\pi_i^k\}\subset BR_i(\boldsymbol{\pi}_{-i})$ with $\pi_i^k\to\pi_i$. Since every $\pi_i^k$ attains the minimum value of $\bar J_i$ and $\bar J_i$ is continuous, the limit $\pi_i\in BR_i(\boldsymbol{\pi}_{-i})$, which shows that $BR_i(\boldsymbol{\pi}_{-i})$ is a closed subset of the compact
set $\Pi_i$, and hence is compact. Moreover, by the quasi convexity of $\bar J_i$ with respect to $\pi_i$, for any
$\pi_i^1,\pi_i^2\in BR_i(\boldsymbol{\pi}_{-i})$ and any $\lambda\in[0,1]$, the mixed policy $\pi_i^{\mathrm{mix}}=\lambda\pi_i^1+(1-\lambda)\pi_i^2$
belongs to $\Pi_i$ and has an objective value no larger than the maximum
objective value of $\pi_i^1$ and $\pi_i^2$, which shows that $\pi_i^{\mathrm{mix}}\in BR_i(\boldsymbol{\pi}_{-i})$, and then $BR_i(\boldsymbol{\pi}_{-i})$ is convex.

\textit{Upper hemicontinuity.}
Consider any sequence $\boldsymbol{\pi}_{-i}^k\to\boldsymbol{\pi}_{-i}$ and any sequence $\pi_i^k\in BR_i(\boldsymbol{\pi}_{-i}^k)$ with $\pi_i^k\to\pi_i$. By the definition of best response, for every $\pi_i'\in\Pi_i$ and every $k$,
\[
\mathbb{E}_{\mathbf z}
\left[
J_i(\pi_i^k,\boldsymbol{\pi}_{-i}^k,z_i,\mathbf z_{-i})
\right]
\le
\mathbb{E}_{\mathbf z}
\left[
J_i(\pi_i',\boldsymbol{\pi}_{-i}^k,z_i,\mathbf z_{-i})
\right].
\]
Taking the limit as $k\to\infty$ and by the continuity of the expected cost to the joint policy profile, we obtain
\[
\mathbb{E}_{\mathbf z}
\left[
J_i(\pi_i,\boldsymbol{\pi}_{-i},z_i,\mathbf z_{-i})
\right]
\le
\mathbb{E}_{\mathbf z}
\left[
J_i(\pi_i',\boldsymbol{\pi}_{-i},z_i,\mathbf z_{-i})
\right],
\]
for every $\pi_i'\in\Pi_i$. Hence, $\pi_i\in BR_i(\boldsymbol{\pi}_{-i})$, which proves that $BR_i$ is upper hemicontinuous.

Therefore, for each slice agent $i$, the best response correspondence $BR_i$ is compact valued, convex valued, and upper hemicontinuous. Consequently, the joint best response correspondence $BR(\boldsymbol{\pi})= BR_1(\boldsymbol{\pi}_{-1})\times\cdots\times BR_N(\boldsymbol{\pi}_{-N})$ also satisfies these properties.

\textit{Step 4: Apply Glicksberg's Fixed Point Theorem.} By Glicksberg's fixed point theorem: there exists a policy profile $\boldsymbol{\pi}^*\in\Pi$ such that $\boldsymbol{\pi}^*\in BR(\boldsymbol{\pi}^*).$ In other words,
\[
\pi_i^*
\in
\arg\min_{\pi_i\in\Pi_i}
\mathbb{E}_{\mathbf z_{-i}}
[
J_i(\pi_i,\boldsymbol{\pi}_{-i},z_i,\mathbf z_{-i})
],\quad \forall i
\]
which is exactly the no unilateral policy improvement condition in the definition of PCE.
\end{proof}

\subsection{Convergence Analysis of Preference Correlated Equilibrium}
\label{subsec:pce_convergence}
We now analyze the convergence of CoSlice toward equilibrium under the regret minimization framework~\cite{regret_minimization}. To accommodate our PCE setting, we define each agent's regret as the cumulative cost relative to its best response. After $K$ policy iterations, the regret of agent $i$ is
\[
R_i(K)
=
\mathbb{E}
\left[
\sum_{k=1}^{K}
\left(
V_i^{\boldsymbol{\pi}^k}(s)
-
V_i^*(s)
\right)
\right],
\]
where $\boldsymbol{\pi}^k$ denotes the joint policy profile, $s$ is the initial state and $V_i^*(s)$ is the minimum expected cost achievable unilaterally. The expectation in each value function is taken over the environment transitions, joint actions, and the correlated coordination signals generated. Accordingly, the total regret of all agents is defined as \(R(K)=\sum_{i\in\mathcal{N}} R_i(K).\)

To derive an upper bound on $R(K)$, we decompose the value gap into two sources of approximation error. The first is the critic estimation error $|Q_i^k(s,\mathbf a,\mathbf z)-Q_i^*(s,\mathbf a,\mathbf z)|$. The second is the actor optimization error $|Q_i^*(s,\mathbf a_i^*,\mathbf a_{-i},\mathbf z) - Q_i^*(s,\mathbf a_i^k,\mathbf a_{-i},\mathbf z)|$. We next bound these two error terms by stochastic optimization analysis.

\begin{lemma}[Critic and Actor Approximation Error Bounds]
\label{lemma:approx_error_bound}
For every agent $i\in\mathcal{N}$, let the critic update follow the recursion
\begin{equation}
\label{eq:critic_update_recursion}
\setlength{\abovedisplayskip}{1pt}
\setlength{\belowdisplayskip}{1pt}
Q_i^{k+1}
=
Q_i^k
+
\eta_k
\left[
\ell_{i,k}
+
\gamma
Q_i^k(s_{k+1},\mathbf a_{k+1},\mathbf z_{k+1})
-
Q_i^k
\right]
\end{equation}
toward the fixed point of the $\gamma$-contracting Bellman operator
\begin{equation}
\label{eq:bellman_operator}
\mathcal T_i Q_i^k
=
\mathbb E
\left[
\ell_{i,k}
+
\gamma
Q_i^k(s_{k+1},\mathbf a_{k+1},\mathbf z_{k+1})
\right],
\end{equation}
and let the actor at iteration $k$ minimize
\begin{equation}
\label{eq:actor_objective}
\setlength{\abovedisplayskip}{1pt}
\setlength{\belowdisplayskip}{1pt}
J_i(\theta_i)
=
\mathbb E
\left[
Q_i^k(s,a_i,\mathbf a_{-i},\mathbf z)
\right],
\qquad
a_i\sim\pi_{\theta_i}(\cdot\mid o_i,z_i),
\end{equation}
by stochastic gradient descent. Suppose (i) the TD target and the stochastic policy gradient both have uniformly bounded variance; (ii) the critic step size $\eta_k=\mathcal O(1/k)$ satisfies the Robbins--Monro conditions~\cite{borkar}; and (iii) $J_i(\theta_i)$ is convex and smooth over a compact parameter space, with actor step size $\alpha_k=\mathcal O(1/\sqrt{k})$. Then the critic estimation error and the actor optimization error,
\begin{equation}
\label{eq:critic_error_definition}
\setlength{\abovedisplayskip}{1pt}
\setlength{\belowdisplayskip}{1pt}
\epsilon_k
:=
\max_{i\in\mathcal N}
\mathbb E
\left[
\left\|
Q_i^k-Q_i^\star
\right\|
\right]
\leq
\frac{C_\epsilon}{\sqrt{k}},
\end{equation}
\begin{equation}
\label{eq:actor_error_definition}
\setlength{\abovedisplayskip}{1pt}
\setlength{\belowdisplayskip}{1pt}
\delta_k
:=
\max_{i\in\mathcal N}
\mathbb E
\left[
Q_i^k(s,a_i^k,\mathbf a_{-i},\mathbf z)
-
Q_i^k(s,a_i^\star,\mathbf a_{-i},\mathbf z)
\right]
\leq
\frac{C_\delta}{\sqrt{k}},
\end{equation}
where $a_i^k\sim\pi_i^k(\cdot\mid o_i,z_i)$ and $a_i^\star\sim\pi_i^\star(\cdot\mid o_i,z_i)$ is sampled from the best unilateral response $\pi_i^\star$ to $\boldsymbol{\pi}_{-i}^k$, hold for constants $C_\epsilon,C_\delta$ independent of $k$.
\end{lemma}

\begin{proof}[Proof of Lemma \ref{lemma:approx_error_bound}]
\textit{Critic.} Since $Q_i^\star=\mathcal T_iQ_i^\star$ and $\mathcal T_i$ is a $\gamma$-contraction, subtracting $Q_i^\star$ from Eq.~\eqref{eq:critic_update_recursion} and using that the TD-error term is zero-mean with bounded variance $\sigma_i^2$ gives
$\mathbb E[\|Q_i^{k+1}-Q_i^\star\|^2] \le (1-(1-\gamma)\eta_k)^2\,\mathbb E[\|Q_i^k-Q_i^\star\|^2] + \eta_k^2\sigma_i^2$.
With $\eta_k=\mathcal O(1/k)$ satisfying the Robbins--Monro conditions, this is a standard stochastic-approximation recursion for a contracting operator~\cite{borkar}, which gives $\mathbb E[\|Q_i^k-Q_i^\star\|^2]\le C_{\epsilon,i}/k$; Jensen's inequality then yields Eq.~\eqref{eq:critic_error_definition} with $C_\epsilon=\max_{i\in\mathcal N}\sqrt{C_{\epsilon,i}}$.

\textit{Actor.} At each iteration, agent $i$ minimizes the convex, smooth objective in Eq.~\eqref{eq:actor_objective} over a compact parameter space using unbiased stochastic gradients of bounded variance and step size $\alpha_k=\mathcal O(1/\sqrt{k})$. The standard convergence rate for stochastic convex optimization~\cite{sgd} directly yields Eq.~\eqref{eq:actor_error_definition} with $C_\delta=\max_{i\in\mathcal N}C_{\delta,i}$.
\end{proof}
We now present our main theoretical result, which establishes a sublinear upper bound on the total regret.

\begin{theorem}[Regret Bound with Approximation Errors]
\label{theo:regret_bound}
Assume the condition of lemma~\ref{lemma:approx_error_bound} hold. Then the total regret satisfies
\[
R(K)
\leq
\frac{N}{1-\gamma}
\sum_{k=1}^{K}
\left(
2\epsilon_k+\delta_k
\right)
=
\mathcal{O}\left(
\frac{N\sqrt{K}}{1-\gamma}
\right).
\]
\end{theorem}

\begin{proof}[Proof of theorem \ref{theo:regret_bound}]
We derive the regret bound using the performance difference lemma~\cite{policy_diff_lemma}, which converts the value gap between two policies into the expected advantage under their induced discounted state distribution.

\textit{Step 1: Express the value gap through Q-value differences.}
At policy iteration $k$, let $\boldsymbol{\pi}^k=(\pi_i^k,\boldsymbol{\pi}_{-i}^k)$ be the current joint policy profile. For each slice agent $i$, let $\pi_i^*$ denote the best unilateral response to $\boldsymbol{\pi}_{-i}^k$. We measure the per iteration regret of agent $i$ by the value gap \(V_i^{\boldsymbol{\pi}^k}(s)
-
V_i^{\pi_i^*,\boldsymbol{\pi}_{-i}^k}(s).\) By the performance difference lemma, this gap can be bounded by the expected Q-value difference between the action induced by the current policy and the action induced by the best unilateral response. Specifically, for actions $a_i\sim \pi_i^k(\cdot\mid o_i,z_i)$ and $a_i^*\sim \pi_i^*(\cdot\mid o_i,z_i)$, we have
\[
\begin{aligned}
&V_i^{\boldsymbol{\pi}^k}(s)
-
V_i^{\pi_i^*,\boldsymbol{\pi}_{-i}^k}(s) \\
&\quad \leq
\frac{1}{1-\gamma}
\mathbb{E}
\left[
Q_i^{\boldsymbol{\pi}^k}
(s,a_i,\mathbf a_{-i})
-
Q_i^{\boldsymbol{\pi}^k}
(s,a_i^*,\mathbf a_{-i})
\right],
\end{aligned}
\]
where the expectation is taken over the discounted state distribution, the coordination signals, and the actions induced by the corresponding policies.

Adding and subtracting the limiting critic $Q_i^*$ at both actions gives
\[
\begin{aligned}
&Q_i^\star(s,a_i^k,\mathbf a_{-i},\mathbf z)
-
Q_i^\star(s,a_i^\star,\mathbf a_{-i},\mathbf z) \\
={}&
\underbrace{
Q_i^\star(s,a_i^k,\mathbf a_{-i},\mathbf z)
-
Q_i^k(s,a_i^k,\mathbf a_{-i},\mathbf z)
}_{\mathrm{I}} \\
&+
\underbrace{
Q_i^k(s,a_i^k,\mathbf a_{-i},\mathbf z)
-
Q_i^k(s,a_i^\star,\mathbf a_{-i},\mathbf z)
}_{\mathrm{II}} \\
&+
\underbrace{
Q_i^k(s,a_i^\star,\mathbf a_{-i},\mathbf z)
-
Q_i^\star(s,a_i^\star,\mathbf a_{-i},\mathbf z)
}_{\mathrm{III}} .
\end{aligned}
\]

\textit{Step 2: Bound the regret by approximation error.}
By Lemma~\ref{lemma:approx_error_bound}, $\mathrm{I}$ and $\mathrm{III}$ are critic error $\epsilon_k$ bounded by $C_{\epsilon}/\sqrt{k}$, and $\mathrm{II}$ is actor error $\delta_k$ bounded by
$C_{\delta}/\sqrt{k}$. Therefore,
\[
V_i^{\boldsymbol{\pi}^k}(s)
-
V_i^{\pi_i^*,\boldsymbol{\pi}_{-i}^k}(s)
\leq
\frac{2C_{\epsilon}+C_{\delta}}{(1-\gamma)\sqrt{k}}.
\]
Summing over all $N$ slice agents and $K$ policy iterations gives
\[
R(K)
\leq
\frac{N(2C_{\epsilon}+C_{\delta})}{1-\gamma}
\sum_{k=1}^{K}\frac{1}{\sqrt{k}}.
\]
Since $\sum_{k=1}^{K}1/\sqrt{k}\leq 2\sqrt{K}$, we have
\[
R(K)
\leq
\frac{2N(2C_{\epsilon}+C_{\delta})}{1-\gamma}
\sqrt{K}
=
\mathcal{O}\left(
\frac{N\sqrt{K}}{1-\gamma}
\right).
\]
\end{proof}

\section{CoSlice: System Design}
\label{sec:method}
\begin{figure}[t]
    \centering
    \includegraphics[width=\columnwidth]{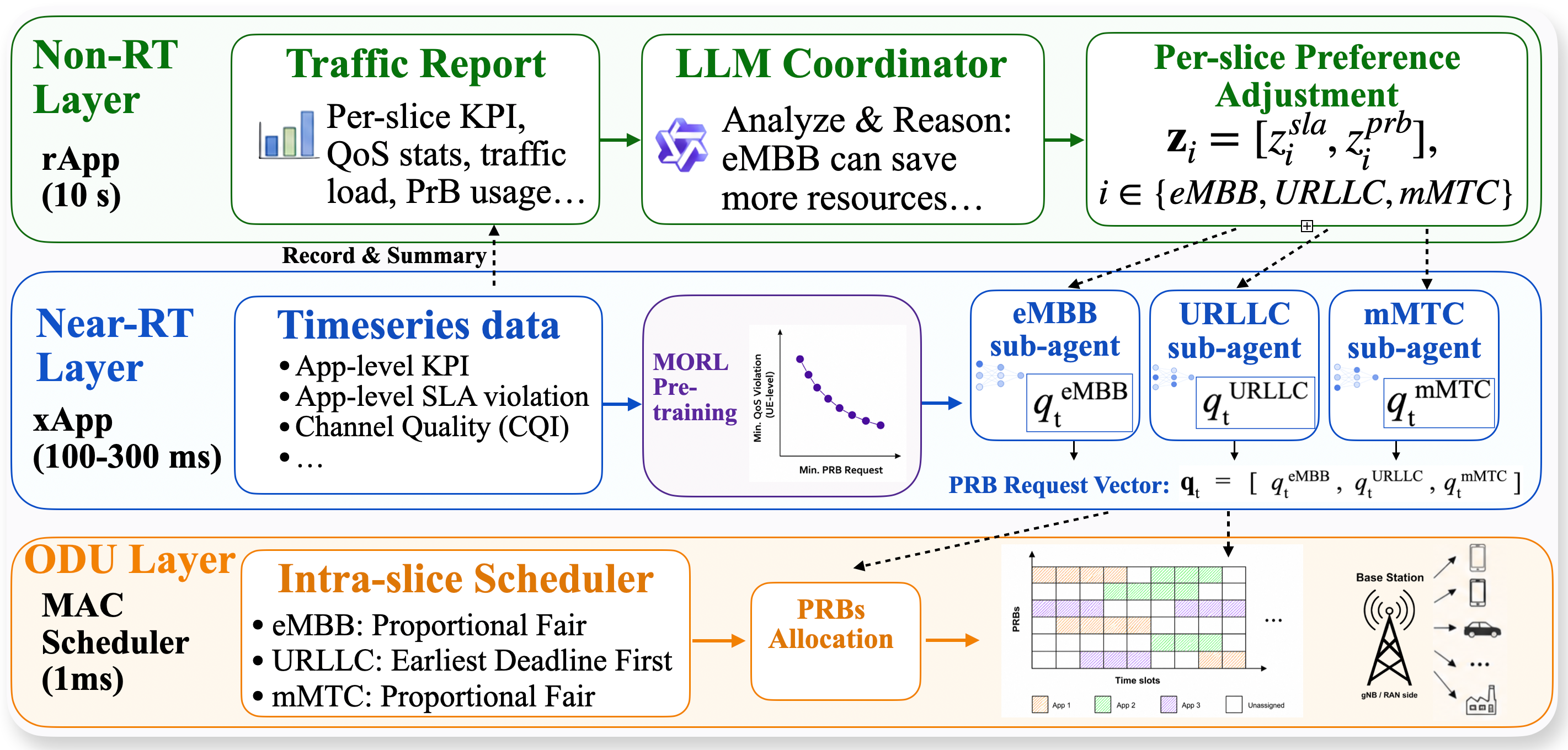}
    \caption{System overview of CoSlice. From top to bottom, the architecture consists of an LLM coordinator in the Non-RT RIC, slice agents in the Near-RT RIC, and a PRB scheduler in the O-DU. Our design focuses on the first two layers.}
    \label{fig:system_overview}
\end{figure}
In this section, we present the system design of CoSlice. It combines an LLM coordinator operating at the Non-RT RIC with preference-conditioned slice agents deployed at the Near-RT RIC, enabling network-wide coordination while preserving fast local resource control. We first introduce the system overview (Sec.~\ref{subsec:system_overview}) and the implementation details of the xApp and rApp controllers (Sec.~\ref{subsec:controller}), followed by the two-stage training pipeline (Sec.~\ref{subsec:training_pipeline}).

\subsection{System Overview}
\label{subsec:system_overview}
CoSlice is organized into three layers operating at different decision timescales, as shown in Fig.~\ref{fig:system_overview}. This design follows the standard O-RAN architecture, where the LLM coordinator is deployed as a Non-RT rApp and the slice agents are deployed as Near-RT xApps, allowing CoSlice to integrate with existing O-RAN interfaces without modifying the underlying control plane. The Non-RT coordinator periodically analyzes summarized network reports and updates the optimization preference for each slice, while the Near-RT slice agents continuously generate slice-specific PRB requests according to the current network state and the received preferences.

The PRB requests from all slice agents are collected into a request vector and forwarded to the O-DU for inter-slice resource allocation. The O-DU then performs millisecond-level MAC scheduling within each slice using the allocated resources. In this way, CoSlice separates slower network-wide coordination from fast local resource control while preserving the standard O-RAN execution pipeline.

\subsection{xApp and rApp Controller Details}
\label{subsec:controller}
\begin{figure}[htbp]
    \centering
    \includegraphics[width=\columnwidth]{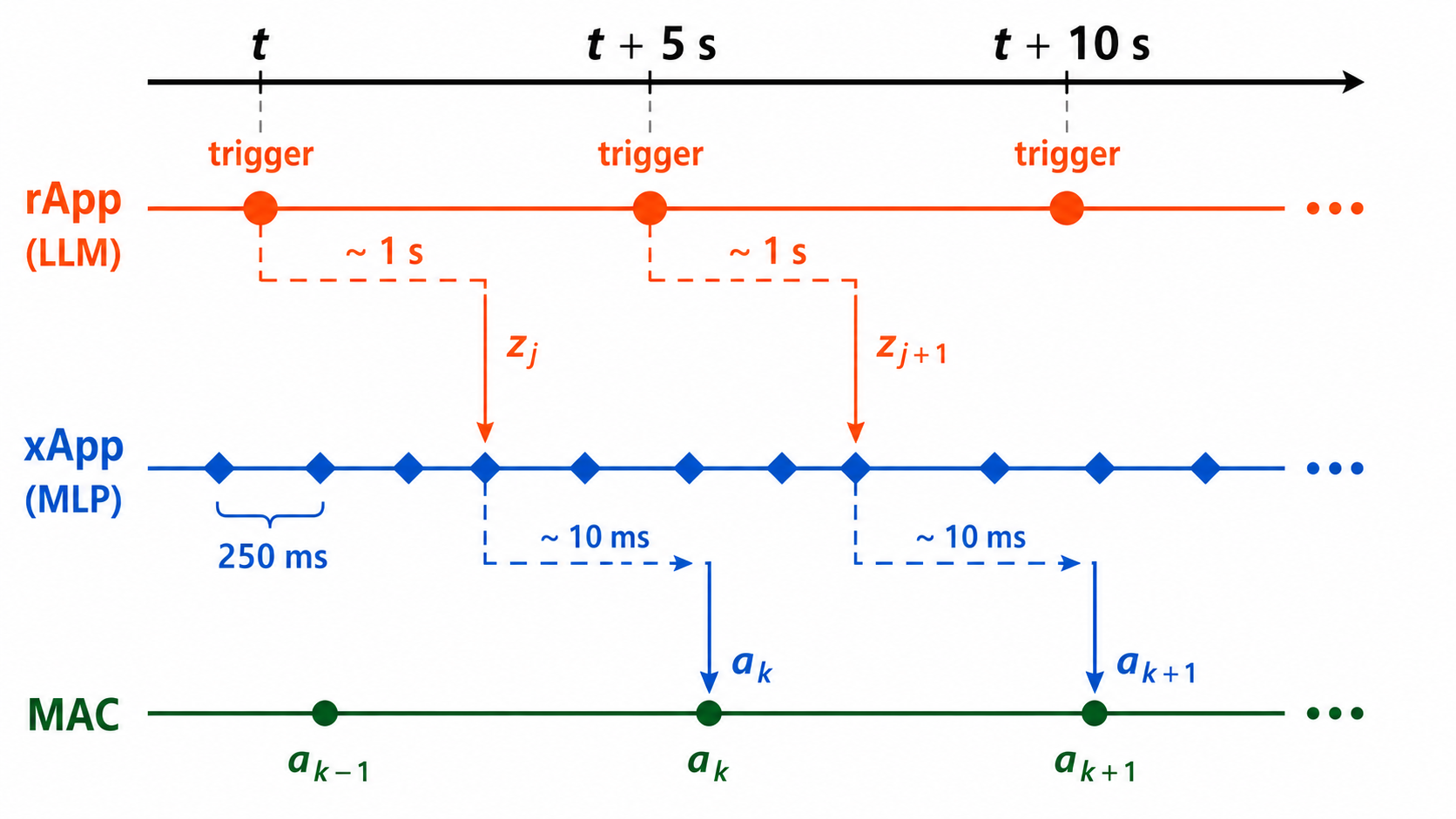}
    \caption{Periodic decisions and delayed control propagation across the rApp, xApp, and MAC layers. Each controller makes a decision from the collected information, but the decision affects the next layer only after inference and message delivery, making low end-to-end latency essential.}
    \label{fig:inference_timeline}
    \vspace{-2mm}
\end{figure}

As illustrated in Fig.~\ref{fig:inference_timeline}, CoSlice executes asynchronously across the Non-RT and Near-RT control loops. The LLM coordinator periodically updates slice preferences, while the slice agents continuously generate PRB requests. As a result, LLM inference and signaling do not block the decision making of the slice agents, while Near-RT inference and signaling do not block MAC scheduling at the O-DU.

At the Near-RT RIC, each slice agent is implemented as a preference-conditioned stochastic policy. Each actor is a multilayer perceptron (MLP) with two fully connected hidden layers of 32 neurons and ReLU activations. The input consists of traffic demand, QoS measurements, resource utilization, recent temporal changes, and the preference vector generated by the LLM coordinator, which is concatenated with the observation, following preference-conditioned function approximation in MORL community~\cite{conditional_network}. QoS metrics, including throughput, latency, and reliability, are normalized by slice-specific SLA targets, so that deviations from the SLA are represented on a comparable scale across different slice types. The actor outputs the mean and standard deviation of a Gaussian action distribution, from which an action is sampled and converted into a PRB request. The requests are applied directly whenever the aggregate demand is within the available budget; otherwise, they are proportionally normalized to ensure feasible resource allocation.

At the Non-RT RIC, the coordinator is deployed as a persistent vLLM server~\cite{vLLM}, keeping the model resident in GPU memory across coordination rounds to reduce inference latency. During each coordination period, the xApp summarizes recent network measurements into a compact report containing aggregated traffic statistics, QoS metrics, SLA violations, PRB utilization, and the current slice preferences. Based on this report, the LLM generates updated preference vectors for all slices in a structured JSON format. To ensure format compliance, we fine-tune the model for structured output generation and fall back to the previous preference whenever the generated response cannot be parsed. To improve control stability, we limit the maximum preference adjustment in each coordination round.

\subsection{Training Pipeline}
\label{subsec:training_pipeline}
CoSlice adopts a two-stage training pipeline summarized in Alg.~\ref{alg:coslice_training}. It first learns a preference-conditioned policy through multi-objective reinforcement learning (MORL)~\cite{MODRL}, and then jointly fine-tunes them under the centralized-training and decentralized-execution (CTDE)~\cite{MADDPG} framework to capture cross-slice interactions. During execution, each actor requires only its local observation and preference.

\paragraph{Stage I: Preference-conditioned MORL pretraining.}
Instead of training a separate policy for every preference, CoSlice learns a preference-conditioned policy by sampling preference vectors from $\mathcal{W}$ during MORL pretraining. The slice-level objective is converted into a scalar penalty through linear scalarization $\ell_t^i=z_{\mathrm{sla}}^i\ell_{\mathrm{sla}}^i+ z_{\mathrm{prb}}^i\ell_{\mathrm{prb}}^i$. In practice, the sampled preferences avoid extreme weightings (e.g.,$[1,0]$ and $[0,1]$) by a small margin, since such extreme preferences are rarely useful in practical network slicing. The preference-conditioned policy is then trained with a PPO-style clipped objective~\cite{schulman2017proximal}. After pretraining, the policy learns to directly respond to the preference signal, enabling each slice agent to adapt its resource allocation behavior according to different SLA-versus-PRB priorities.

\begin{figure}[htbp]
    \centering
    \includegraphics[width=\columnwidth]{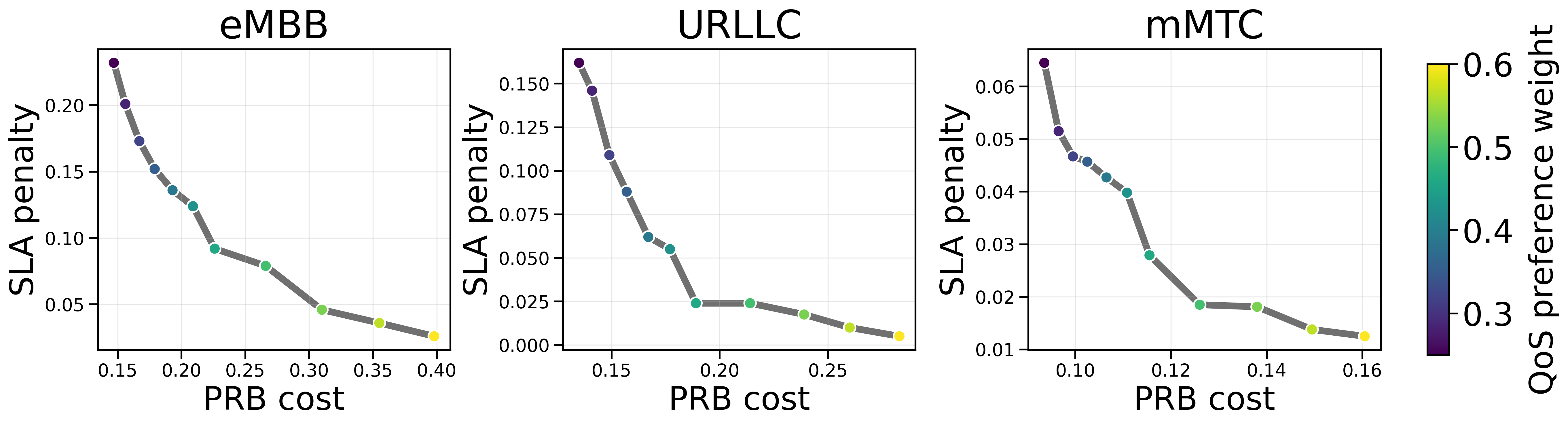}
    \caption{Pareto front of each slice approximated by MORL pretraining. Each point corresponds to an operating point under a different SLA-versus-PRB preference. A larger SLA weight leads to lower SLA violation but generally requires more PRBs, whereas a larger PRB-saving weight favors more conservative requests. The resulting policy can represent multiple Pareto-optimal solutions, providing a suitable initialization for subsequent multi-agent training.}
    \label{fig:pareto_front}
\end{figure}

\paragraph{Stage II: Joint multi-agent training.}
Stage I learns the preference-conditioned behavior of each slice independently, but it does not capture cross-slice interactions under the shared PRB budget. Stage II therefore jointly fine-tunes the pretrained slice policies to improve their coordination. Because network traces form a continuing process without clear episode boundaries, CoSlice updates its policies from step-level transitions instead of complete trajectories. We build the training framework upon TD3~\cite{td3}, a widely adopted actor-critic algorithm based on temporal-difference learning. Specifically, each slice is associated with twin centralized action-value critics $Q_{\phi_{i,1}}^i(\mathbf{s}_t,\mathbf{a}_t)$ and $Q_{\phi_{i,2}}^i(\mathbf{s}_t,\mathbf{a}_t)$, which can improve training stability by reducing value overestimation. The critics are
optimized by minimizing the temporal-difference (TD) loss
\begin{equation}
\label{eq:coslice_critic_loss}
\mathcal{L}_{\mathrm{critic}}
=
\sum_{i=1}^{N}
\mathbb{E}_{t}
\left[
\sum_{m=1}^{2}
\left(
Q_{\phi_{i,m}}^i(\mathbf{s}_t,\mathbf{a}_t)-y_t^i
\right)^2
\right].
\end{equation}
where $y_t^i=\ell_t^i+\gamma(1-d_t)\min_{m\in\{1,2\}}
Q_{\bar{\phi}_{i,m}}^i(\mathbf{s}_{t+1},\mathbf{a}'_{t+1})$ is the TD target, with $\mathbf{a}'_{t+1}$ generated by the target actors. The learned action-value estimates are then used to optimize the stochastic actors. To improve policy stability under rapidly changing
and bursty network traffic, CoSlice applies a PPO-style clipping mechanism to the actor objective:
\begin{equation}
\label{eq:coslice_actor_loss}
\begin{aligned}
\mathcal{L}_{\mathrm{actor}}^i
=
-\mathbb{E}_{t}
\Big[
\min\big(
&\rho_t^iQ_{\phi_i}^i(\mathbf{s}_t,\mathbf{a}_t),\\
&
\operatorname{clip}(\rho_t^i,1\pm\epsilon)
Q_{\phi_i}^i(\mathbf{s}_t,\mathbf{a}_t)
\big)
\Big].
\end{aligned}
\end{equation}
where
$\rho_t^i=
\pi_{\theta_i}(a_t^i\mid\mathbf{o}_t^i,\mathbf{z}_t^i)/
\pi_{\theta_i^{\mathrm{old}}}
(a_t^i\mid\mathbf{o}_t^i,\mathbf{z}_t^i)$
is the policy ratio.

\begin{algorithm}[t]
\caption{Two-stage training procedure of CoSlice.}
\label{alg:coslice_training}
\footnotesize
\begin{algorithmic}[1]
\Require Slice set $\mathcal{I}$; preference domain $\mathcal{W}$; LLM coordinator $\mathcal{C}$

\Statex
\State \textbf{Stage I: MORL pretraining}
\For{each slice $i\in\mathcal{I}$}
    \State Initialize actor $\pi_{\theta_i}$ and critic $V_{\psi_i}$
    \For{each pretraining trajectory}
        \State Sample $\mathbf{z}^i\sim\mathcal{W}$
        \State Collect single-slice transitions using $\pi_{\theta_i}$
        \State Compute the scalarized slice penalty $\ell^i=z_{\mathrm{sla}}^i\ell_{\mathrm{sla}}^i+z_{\mathrm{prb}}^i\ell_{\mathrm{prb}}^i$
        \State Update $(\theta_i,\psi_i)$ using the actor objective and local critic loss
    \EndFor
\EndFor

\Statex
\State \textbf{Stage II: Multi-agent joint training}
\State Initialize pretrained actors and twin centralized critics, and collect the initial monitoring report $\mathcal R$
\For{each joint-training interval $k$}
    \State
    $\{\mathbf{z}_k^i\}_{i=1}^{N}\gets\mathcal{C}(\mathcal{R})$
    \For{each decision step $t=1,\ldots,T$}
        \For{each slice $i\in\mathcal{I}$ \textbf{in parallel}}
            \State Sample
            $a_t^i\sim
            \pi_{\theta_i}(\cdot\mid\mathbf{o}_t^i,\mathbf{z}_k^i)$
        \EndFor
        \State Execute normalized $\mathbf{a}_t$ at the O-DU
        \State Observe network measurements and slice penalties
        \State Store $(\mathbf{s}_t,\mathbf{a}_t,\ell_t, \mathbf{s}_{t+1},d_t)$ in the replay buffer
        \State Sample a mini-batch and update the twin critics using Eq.~\eqref{eq:coslice_critic_loss}
        \State Soft-update the target critics
        \If{the delayed actor update is triggered}
            \State Update the actor policies $\{\theta_i\}_{i=1}^{N}$ using Eq.~\eqref{eq:coslice_actor_loss}
            \State Soft-update the target actors
        \EndIf
        \State Refresh monitoring report $\mathcal{R}$
    \EndFor
\EndFor
\State \Return $\{\pi_{\theta_i}\}_{i=1}^{N}$
\end{algorithmic}
\end{algorithm}

\section{Experiment}
\label{sec:experiment}
In this section, we evaluate the proposed CoSlice framework on the ORANSlice/OAI RFsim testbed. We first describe the experimental setup, including the emulation environment, baselines, and evaluation metrics in Subsec.~\ref{subsec:exp_setting}. We then evaluate CoSlice under different slicing settings in Subsec.~\ref{subsec:performance_analysis}, analyze the contribution of the LLM coordinator through case studies and ablation experiments in Subsec.~\ref{subsec:llm_demo}, and finally measure the end-to-end inference latency in Subsec.~\ref{subsec:inference_time}. The source code will be made publicly available after the review process.

\subsection{Experimental Settings}
\label{subsec:exp_setting}
\paragraph{ORAN-Slice + RFsim emulator} We build the network environment using ORANSlice~\cite{oranslice}, which extends the OpenAirInterface (OAI) protocol stack with O-RAN compatible slicing control. Unlike purely Python-based simulators that derive network performance from analytical traffic and capacity models, our environment executes a software-based 5G system comprising the OAI 5G Core, an ORANSlice-enabled gNB, and multiple independent OAI New Radio User Equipment (nrUE) instances. All components are deployed in isolated Docker containers with independent processes, network interfaces, and protocol states, thereby emulating the separation among physical network entities. Moreover, the OAI gNB mirrors the protocol-stack architecture and packet-processing pipeline of a real 5G gNB; traffic from different slices is identified by its S-NSSAI and processed through the corresponding SDAP/PDCP, RLC, and slice-aware MAC scheduling paths before transmission. Furthermore, To interconnect the software gNB and UEs, we use the OAI RF simulator (RFsim)~\cite{oai_rfsim}, which replaces the physical radio hardware and over-the-air channel by transporting time-domain baseband samples between the gNB and nrUEs. Slice-control actions are enforced by the ORANSlice MAC scheduler as per-slice PRB policies, and the resulting throughput, delay, jitter, and packet loss are obtained from actual user-plane traffic and RAN measurements.

\paragraph{Performance Baselines}
In addition to CoSlice, we implement the following inter-slice resource allocation policies in ORANSlice for comparison.
\begin{itemize}
\item \textbf{Equal:}
A static policy that equally divides the available physical resource blocks (PRBs) among all slices throughout the experiment. It serves as a simple reference baseline.

\item \textbf{Zipper~\cite{zipper}:}
Zipper formulates network slicing as a model predictive control (MPC) problem and deploys an independent DNN controller for each slice. At runtime, the controllers estimate resource demands, and use dynamic programming to resolve resource contention under the PRB budget.

\item \textbf{xSlice~\cite{xslice}:}
xSlice implements an O-RAN xApp using reinforcement learning with a reward balancing SLA satisfaction and PRB utilization. It employs a graph neural network (GNN) to aggregate user-level observations and jointly allocates resources across all slices.

\item \textbf{COQRA~\cite{COQRA}:}
COQRA deploys an independent Q-learning agent for each slice. The agents exchange Q-values and coordinate resource allocation by computing a correlated equilibrium under the shared PRB constraint.
\end{itemize}

\paragraph{Evaluation Metrics and QoS Measurement}
We use the relative SLA violation in Eq.~\ref{eq:relative_sla_penalty} as the primary evaluation metric. For each slice, we report the mean, 95-th, and 99-th percentile violations to evaluate both average and tail performance. Since slices have different SLA requirements, results are reported separately for each slice, together with the macro-average of the mean violations. Specifically, eMBB requires a minimum downlink throughput of \(3.5\,\mathrm{Mbps}\) per active user, URLLC requires latency and jitter below \(5\,\mathrm{ms}\) with packet loss below \(0.5\%\), and mMTC requires at least \(90\%\) event delivery with latency below \(20\,\mathrm{ms}\) and packet loss below \(10\%\). All QoS metrics are measured from actual packet transmissions during each control interval, excluding inactive users. The downlink throughput is computed as $T_{i,t}^{\mathrm{DL}} = 8B_{i,t} / (\Delta t_{i,t} \times 10^{6})$, where $B_{i,t}$ is the received bytes and $\Delta t_{i,t}$ is the measurement duration. Packet latency and loss are measured using sequence numbers and transmission timestamps carried in each UDP packet. All methods are evaluated under identical traffic seeds, environment settings, initial resource configurations, and a 5\,s warm-up period.

\subsection{Performance Analysis under Different Slice Settings}
\label{subsec:performance_analysis}

\begin{figure}[t]
    \centering
    \begin{subfigure}[t]{\columnwidth}
        \centering
        \includegraphics[width=\linewidth]{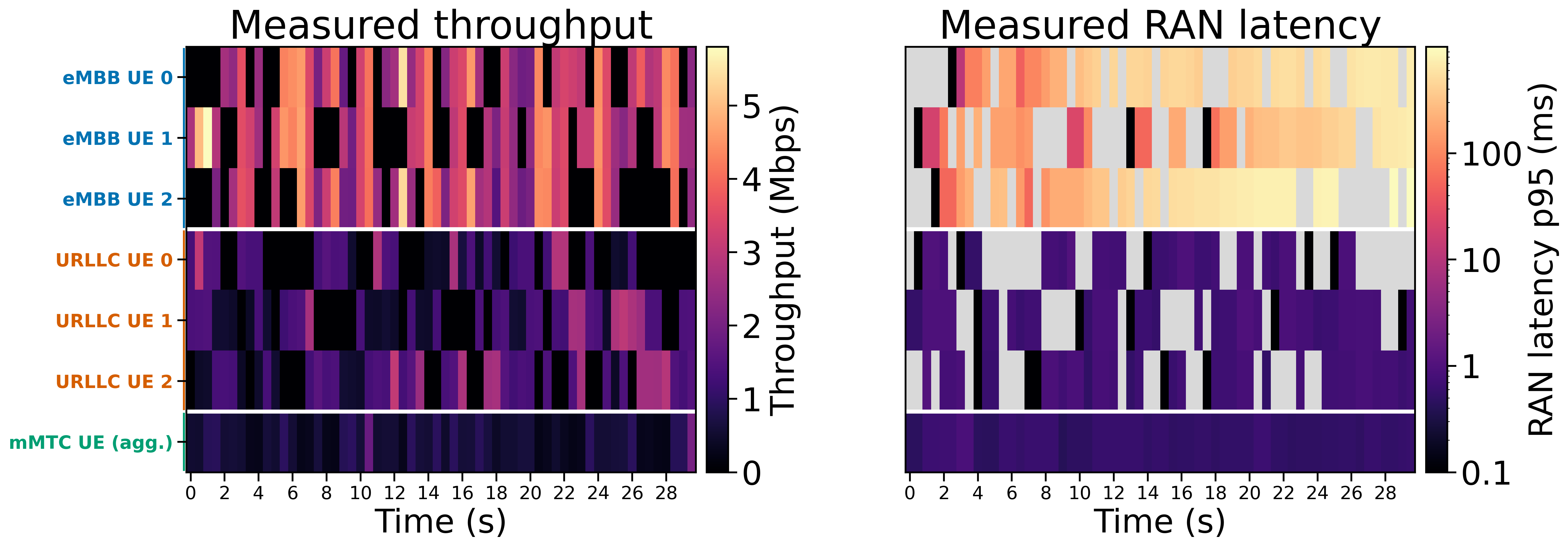}
        \caption{3-slice setting.}
        \label{fig:traffic_trace_3slice}
    \end{subfigure}

    \begin{subfigure}[t]{\columnwidth}
        \centering
        \includegraphics[width=\linewidth]{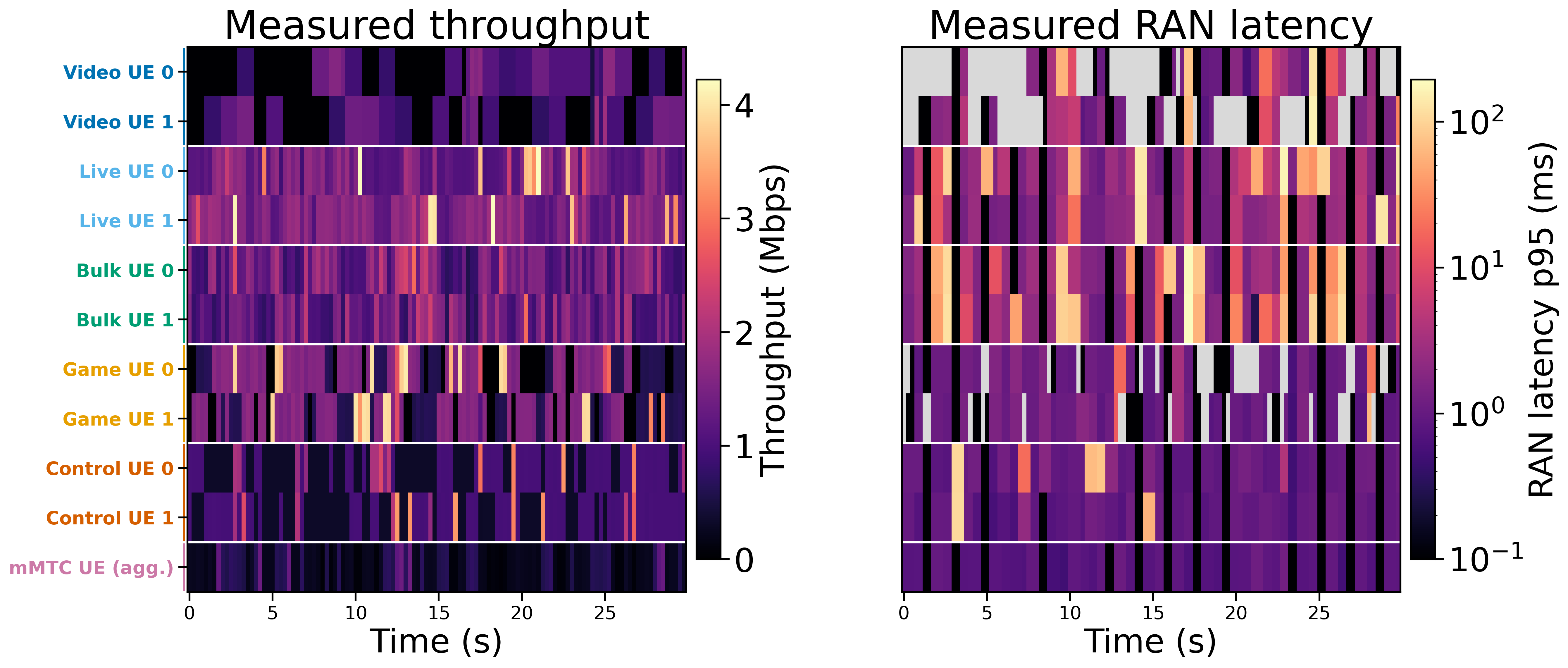}
        \caption{6-slice setting.}
        \label{fig:traffic_trace_6slice}
    \end{subfigure}

    \caption{Measured throughput and RAN latency trace data in the three- and six-slice settings. The traces show distinct service dynamics: eMBB exhibits high-rate bursty or sustained transmissions, whereas URLLC shows intermittent latency-sensitive activity and mMTC presents lower-rate aggregated demand. The six-slice setting further exposes differentiated application-level patterns within these service classes.}
    \label{fig:traffic_traces}
    \vspace{-2mm}
\end{figure}

\paragraph{Three-slice setting}
We first consider a three-slice setting comprising eMBB, URLLC, and mMTC services, whose measured traffic traces are shown in Fig.~\ref{fig:traffic_trace_3slice}. Each UE belongs to a single slice and runs an independent application workload. The eMBB slice contains four UEs emulating buffered DASH video clients, while the URLLC slice contains five UEs running an interactive gaming application over custom UDP traffic. Since directly emulating hundreds of IoT devices with OAI nrUE instances is impractical, the mMTC slice is modeled as an aggregate event stream. During each traffic generation interval \(T\), the number of events follows a Poisson distribution with rate $\lambda$, and event arrivals are uniformly distributed within the interval. Traffic intensity evolves according to a three-state Markov model (low, medium, and high load), with occasional Poisson traffic bursts.

\begin{figure}[t]
    \centering

    \makebox[\linewidth][c]{%
        \hspace*{0.04\linewidth}%
        \raisebox{0.0mm}{%
            \includegraphics[width=0.92\linewidth]
                {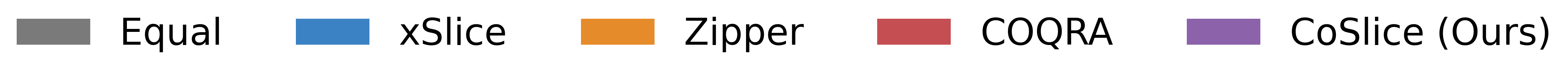}%
        }%
    }
    \vspace{-1mm}

    \begin{subfigure}[t]{0.485\linewidth}
        \centering
        \includegraphics[width=\linewidth]
            {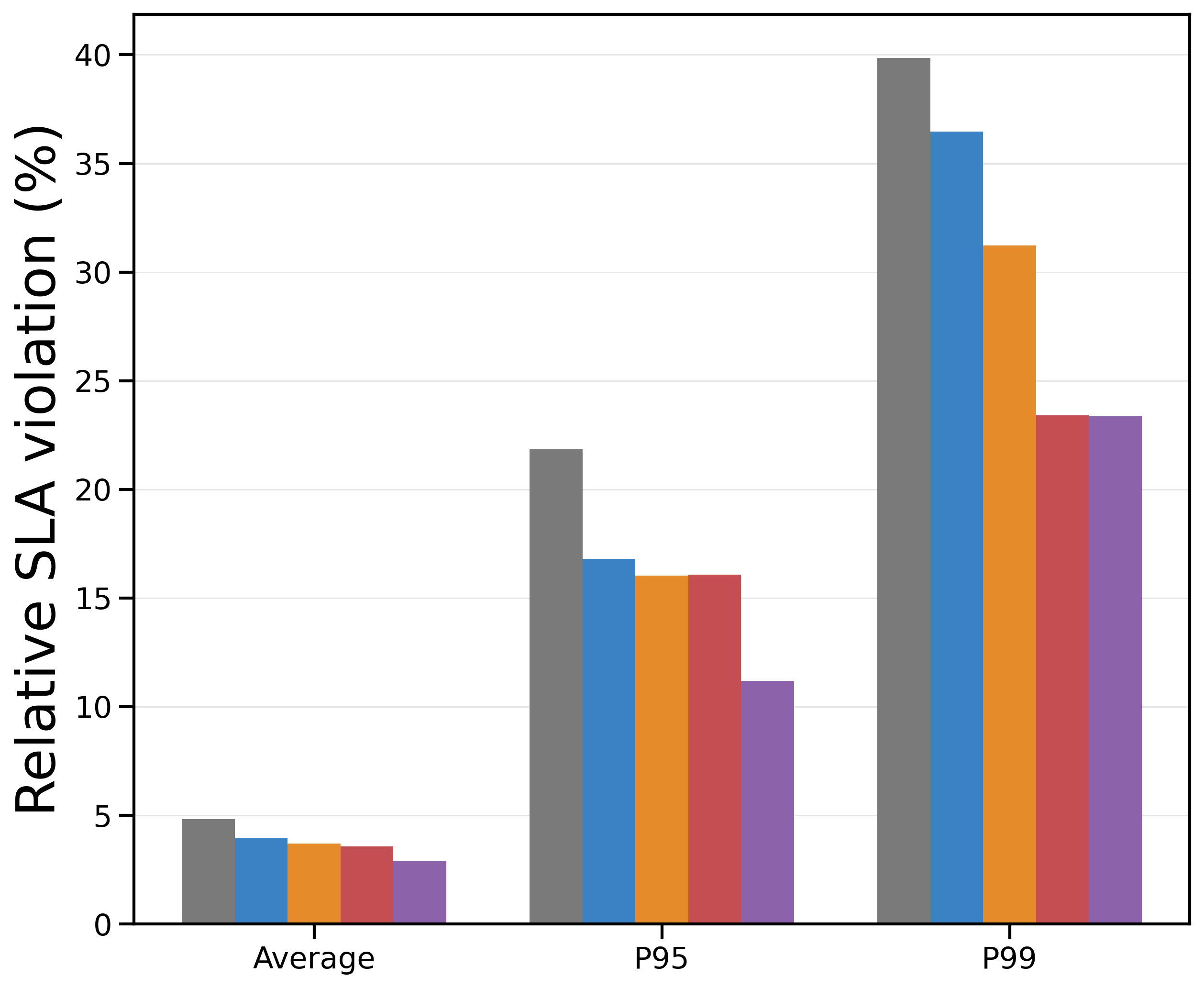}
        \caption{3-slice setting.}
        \label{fig:macro_sla_3slice}
    \end{subfigure}
    \hfill
    \begin{subfigure}[t]{0.485\linewidth}
        \centering
        \includegraphics[width=\linewidth]
            {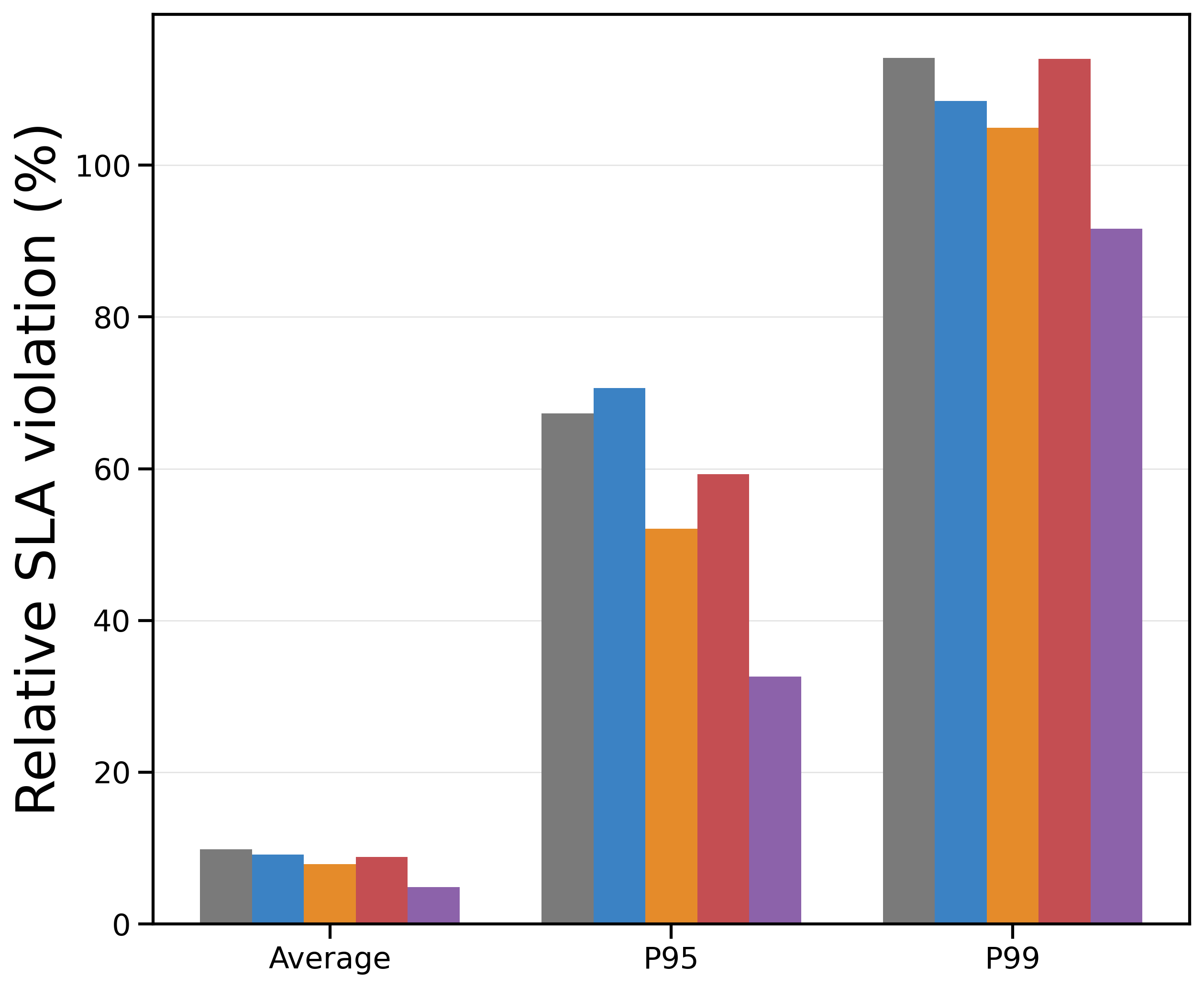}
        \caption{6-slice setting.}
        \label{fig:macro_sla_6slice}
    \end{subfigure}
    \caption{Comparison of all algorithms in the three- and six-slice settings. We report the macro-averaged mean and tail (95th- and 99th-percentile) relative SLA violations across all slices; lower values are better. CoSlice achieves the lowest values for all three statistics in both settings. Compared with the baseline with best performance, CoSlice reduces the mean/P95/P99 violations by 19.4\%/30.2\%/0.2\% in the three-slice setting and 37.9\%/37.4\%/12.7\% in the six-slice setting.}
    \label{fig:relative_sla_violation}
    \vspace{-2mm}
\end{figure}

\begin{figure}[htbp]
    \centering
    \includegraphics[width=\linewidth]
    {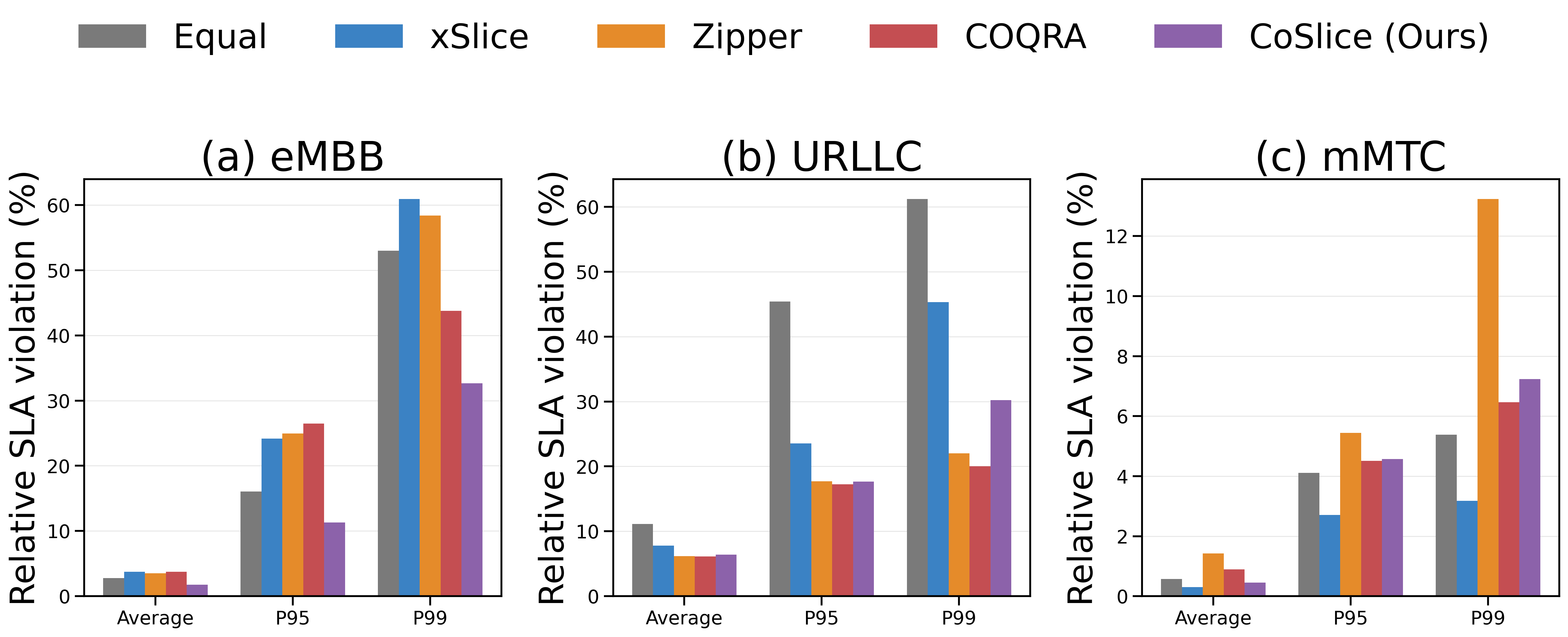}
    \caption{Per-slice SLA performance in the three-slice setting. CoSlice leads consistently on eMBB, with moderate tradeoffs on URLLC and mMTC; the mMTC violations remain small.}
    \label{fig:main_results_3slice}
    \vspace{-2mm}
\end{figure}

Fig.~\ref{fig:relative_sla_violation} compares the macro-averaged SLA performance across all methods, where CoSlice demonstrates the strongest overall system performance. The per-slice results in Fig.~\ref{fig:main_results_3slice} further reveal the performance on all slices. CoSlice substantially improves the eMBB slice, whose bursty traffic creates highly dynamic resource demand, and maintains consistently low violations for the latency-sensitive URLLC slice. While the mMTC slice exhibits slightly higher tail violations, the degradation remains limited and is outweighed by the substantial improvements achieved for the other slices, resulting in the strongest overall SLA performance.

\paragraph{Six-slice setting}
We further evaluate CoSlice in a finer-grained slicing setting, where users within the same service category exhibit different traffic dynamics~\cite{zipper,network_slicing_survey}. Specifically, eMBB is divided into video-on-demand, live-streaming, and bulk-transfer slices, URLLC into interactive-gaming and control-loop slices, while mMTC remains as an aggregated slice, resulting in six slices with diverse application-level traffic patterns (Fig.~\ref{fig:traffic_trace_6slice}). Each slice uses an application-specific workload model matching its traffic characteristics.

\begin{figure}[htbp]
    \centering
    \includegraphics[width=\linewidth]
    {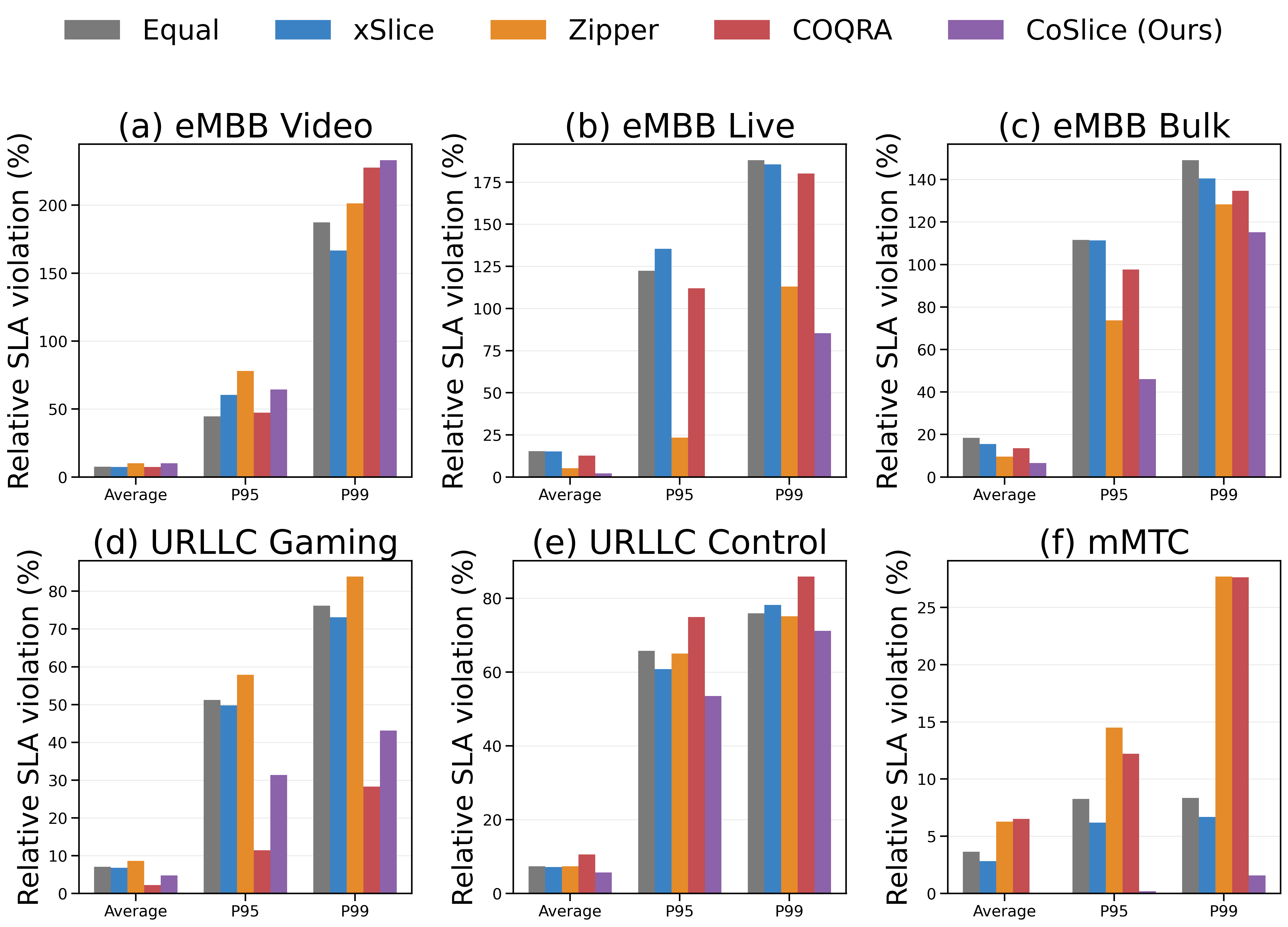}
    \caption{Per-slice SLA performance in the six-slice setting. CoSlice leads on eMBB Live, eMBB Bulk, URLLC Control, and aggregated mMTC, while trading performance on eMBB Video and URLLC Gaming for the strongest overall macro-average result.}
    \label{fig:main_results_6slice}
    \vspace{-2mm}
\end{figure}

Fig.~\ref{fig:relative_sla_violation} shows that CoSlice continues to achieve the strongest macro-averaged SLA performance under this more challenging setting. Specifically, the per-slice results in Fig.~\ref{fig:main_results_6slice} demonstrate that CoSlice consistently improves the slices with the highest resource demand and the strictest latency requirements, while maintaining competitive performance on the remaining slices. As application-level traffic becomes heterogeneous, coordinated resource allocation becomes more important than optimizing slices independently, allowing CoSlice to achieve the best overall system performance.

\subsection{How Does the LLM Signal Improve Coordination?}
\label{subsec:llm_demo}
\paragraph{Case study: interpretable LLM-guided preference adaptation} In our experiments, we deploy Qwen3-4B-Instruct~\cite{qwen3} in the Non-RT RIC, where it periodically analyzes aggregated network statistics from all slices, and adapts their preference signals according to the current network conditions. Benefiting from the rich prior knowledge encoded in pretrained LLMs, CoSlice can reason over diverse network statistics and dynamically adjust slice preferences as network conditions evolve. Moreover, the LLM output provides interpretable textual reasoning, explaining how the observed network statistics lead to each preference update, which is an important capability for trustworthy network management~\cite{inslice}. As shown in Fig.~\ref{fig:llm_case_study}, we present one decision step of the LLM coordinator. The input consists of a system prompt together with a measured traffic report, and the output contains a global \emph{Think} statement, slice-specific \emph{Reason}s, and the corresponding preference updates.

\begin{figure}[t]
  \centering
  \begin{tikzpicture}
    \node[
      draw=LLMInput,
      fill=white,
      line width=0.5pt,
      rounded corners=1mm,
      inner sep=1.5mm
    ] {%
      \begin{minipage}{0.92\columnwidth}
        \scriptsize
        \noindent
        \textcolor{LLMInput}{\textbf{LLM INPUT}}
        \par\smallskip
        \noindent
        \textbf{System Prompt.}
        You are the NonRT-RIC coordinator for CoSlice. Analyze the measured ORAN-Slice/RFsim report below and independently choose one direction for each slice: increase SLA priority, hold the current preference, or
        increase PRB-saving priority. Consider the SLA severity in percentage points, material and severe-event frequencies, dominant violation causes, requested resources, and global capacity competition. Return one concise global Think statement and one evidence-grounded Reason for each decision.

        \par\medskip
        \hrule height 0.45pt
        \par\medskip

        \noindent
        \textbf{Measured Traffic Report.}
        The report contains 100 measured control steps. The current SLA preference is \(z_{\mathrm{SLA}}=0.50\) for all slices \(\ldots\)
        \par\smallskip
        \noindent
        \textcolor{LLMBlue}{\textbf{eMBB:}}
        SLA mean/P95/P99 =
        \(9.224/49.225/80.000\) points;
        material/severe frequency =
        \(0.294/0.252\);
        requested/applied PRB ratio =
        \(0.3517/0.3256\) \(\ldots\)

        \par\smallskip
        \noindent
        \textcolor{LLMOrange}{\textbf{URLLC:}}
        SLA mean/P95/P99 =
        \(14.775/57.584/68.689\) points;
        material/severe frequency =
        \(0.510/0.507\);
        requested/applied PRB ratio =
        \(0.3945/0.3650\) \(\ldots\)

        \par\smallskip
        \noindent
        \textcolor{LLMOutput}{\textbf{mMTC:}}
        SLA mean/P95/P99 =
        \(0.434/2.585/4.042\) points;
        material/severe frequency =
        \(0.160/0.000\);
        requested/applied PRB ratio =
        \(0.2897/0.2684\).
        No severe event is observed. \(\ldots\)

      \end{minipage}%
    };
  \end{tikzpicture}

  \vspace{1.2mm}

  \begin{tikzpicture}
    \node[
      draw=LLMOutput,
      fill=white,
      line width=0.5pt,
      rounded corners=1mm,
      inner sep=1.5mm
    ] {%
      \begin{minipage}{0.92\columnwidth}
        \scriptsize

        \noindent
        \textcolor{LLMOutput}{\textbf{LLM OUTPUT}}
        \par\smallskip

        \noindent
        \textbf{Think.}
        \emph{``eMBB and URLLC face substantial tail risks, while mMTC can safely prioritize PRB saving.''}

        \par\medskip
        \hrule height 0.45pt
        \par\medskip

        \noindent
        \textcolor{LLMBlue}{\textbf{eMBB}} 
        Action: Increase SLA priority. Reason: SLA \(P_{95}\) is \(49.225\) points and the severe-event frequency is \(0.252\), with throughput contributing \(9.063\) points on average. The SLA preference increases from
        \(0.50\rightarrow0.60\).
        
        \par\smallskip
        
        \noindent
        \textcolor{LLMOrange}{\textbf{URLLC}}
        Action: Increase SLA priority. Reason: SLA \(P_{95}\) is \(57.584\) points and the severe-event frequency is \(0.507\), with latency contributing \(12.743\) points on average. The SLA preference increases from \(0.50\rightarrow0.60\).
        
        \par\smallskip
        
        \noindent
        \textcolor{LLMOutput}{\textbf{mMTC}}
        Action: Increase PRB-saving priority. Reason: The mean SLA penalty is \(0.434\) points, SLA \(P_{95}\) is \(2.585\) points, and no severe event is observed. The SLA preference decreases from \(0.50\rightarrow0.40\).
      \end{minipage}%
    };
  \end{tikzpicture}
  \caption{Case study of one LLM decision step.}
  \label{fig:llm_case_study}
\end{figure}

\begin{figure}[htbp]
    \centering
    \includegraphics[width=\columnwidth]{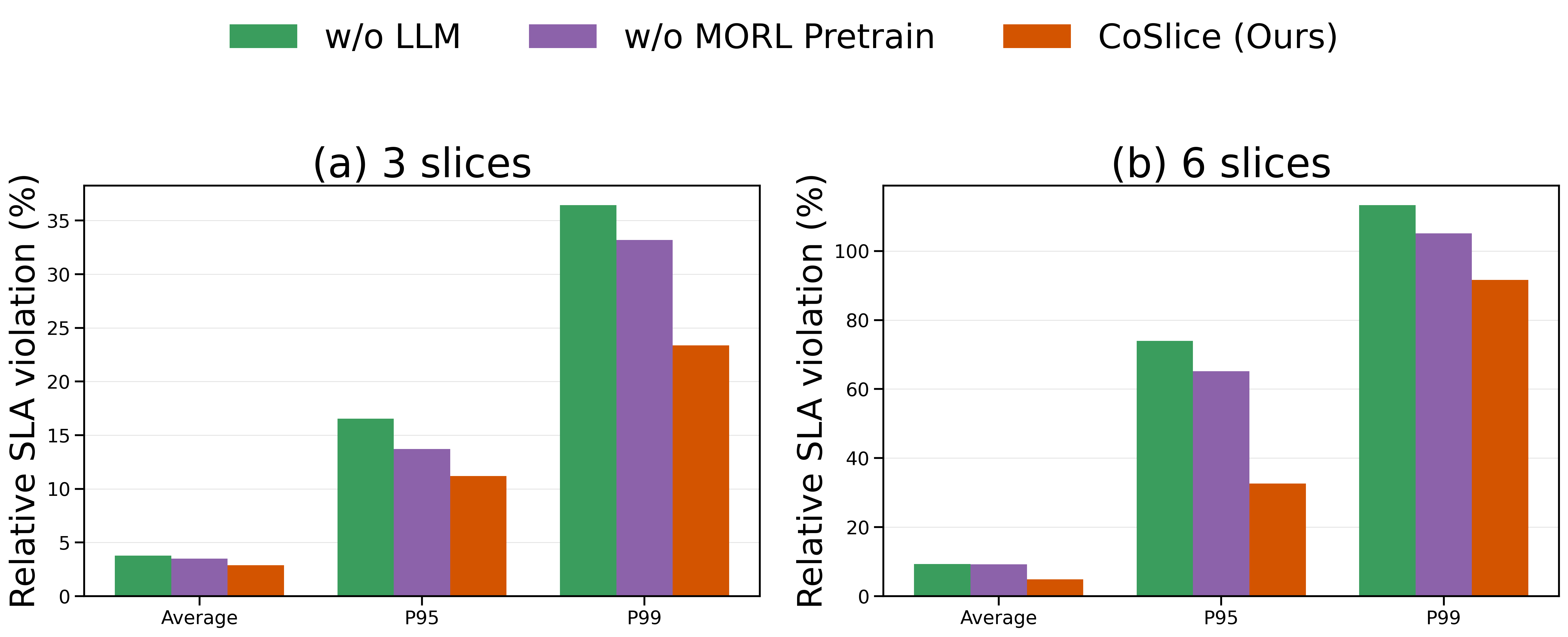}
    \caption{Ablation study of CoSlice in the three- and six-slice settings. The bars show the macro-averaged mean and tail (95th- and 99th-percentile) relative SLA violations across slices; lower values are better. Compared to the base version, adding the LLM coordinator reduces the macro-averaged mean and 95th-/99th-percentile violations by 7.6\%/17.2\%/8.9\% in the three-slice setting and 1.0\%/11.9\%/7.2\% in the six-slice setting. Further adding MORL pretraining yields additional reductions of 17.6\%/18.4\%/29.6\% and 47.0\%/49.9\%/12.8\%, respectively.}
    \label{fig:ablation}
    \vspace{-2mm}
\end{figure}

\paragraph{Ablation study: effect of LLM coordination and MORL pretraining} To quantify the contribution of each component, we compare three CoSlice variants: a baseline with independent slice policies, adding the LLM coordinator, and the complete framework with MORL pretraining. Figure~\ref{fig:ablation} reports the macro-averaged mean, 95th-, and 99th-percentile relative SLA violations; All variants are trained under the same ORAN-Slice/RFsim interactions and evaluated using identical traffic seed.

Compared with the baseline, introducing the LLM coordinator consistently reduces both average and tail SLA violations by dynamically reweighting the importance of SLA violation and PRB usage during training. However, each slice is still represented by a single MLP policy that cannot explicitly adapt to the updated preferences. MORL pretraining equips each slice agent with preference-conditioned policies, allowing the complete CoSlice framework to achieve the lowest SLA violations in both the three-slice and six-slice settings.

\subsection{Inference Time}
\label{subsec:inference_time}
Low control latency is essential for timely AI-driven radio resource management. We measure the end-to-end latency of each control component in the three-slice deployment, with the mean and standard deviation reported in Table~\ref{tab:inference_latency}. Experiments are conducted on a server equipped with an AMD EPYC~9354 CPU and an NVIDIA RTX PRO~6000 Blackwell Max-Q GPU. The Qwen3-4B-Instruct coordinator is served locally by vLLM on the GPU, while the lightweight CoSlice xApp models run on the CPU. We further compare the decentralized CoSlice agents with the centralized xSlice implementation under the same hardware and runtime configuration.
\begin{table}[htbp]
    \centering
    \caption{Control latency across different components. All components remain well below their control periods. Local CoSlice agents achieve lower and more stable inference latency than the single xSlice agent due to their simpler input spaces. The LLM coordinator incurs higher latency from autoregressive decoding, but still averages below 1,s. MLP results are averaged over 1000 samples, and LLM results over 30 warm-up-excluded samples.}
    \label{tab:inference_latency}
    \setlength{\tabcolsep}{5pt}
    \begin{tabular}{lccc}
        \toprule
        & \textbf{xSlice} &
        \textbf{MLP subagents} &
        \textbf{LLM agent} \\
        \midrule
        \textbf{End-to-end latency (ms)}
        & $8.4 \pm 0.8$
        & $6.5 \pm 0.8$
        & $508.2 \pm 8.8$ \\
        \bottomrule
    \end{tabular}
\end{table}

\section{Conclusion}
\label{sec:conclusion}
In this paper, we introduced CoSlice, a preference-conditioned multi-agent reinforcement learning framework for 5G network slicing optimization. CoSlice consists of an LLM coordinator and decentralized slice agents, enabling coordinated inter-slice PRB allocation while preserving fast local decision making. By modeling each slice as a multi-objective learning agent and dynamically adapting its objective preference according to network-wide traffic conditions, CoSlice effectively balances SLA satisfaction and resource efficiency across heterogeneous services. We implemented CoSlice on an O-RAN testbed built upon ORANSlice and OpenAirInterface, and evaluated its performance under realistic traffic scenarios. Experimental results demonstrate that CoSlice consistently achieves the best overall QoS performance compared with existing network slicing approaches, while maintaining low inference latency suitable for near-real-time RAN control.

\clearpage

\end{document}